\documentclass{article}

\usepackage{arxiv}

\usepackage[utf8]{inputenc} 
\usepackage[T1]{fontenc}    
\usepackage{hyperref}       
\usepackage{url}            
\usepackage{booktabs}       
\usepackage{amsfonts}       
\usepackage{nicefrac}       
\usepackage{microtype}      
\usepackage{lipsum}		
\usepackage{graphicx}
\usepackage{natbib}
\usepackage{doi}

\usepackage{graphicx}
\usepackage{booktabs}
\usepackage{multirow}
\usepackage{floatrow}
\usepackage{amsmath}
\usepackage{cleveref}
\usepackage{bbding}
\usepackage{algorithm}
\usepackage{algorithmic}

\title{IceDiff: High Resolution and High-Quality Sea Ice Forecasting with Generative Diffusion Prior}

\author{Jingyi Xu$^{1,*}$, Siwei Tu$^{1,*}$, Weidong Yang$^{1, }$\Envelope, Shuhao Li$^1$, Keyi Liu$^1$, Yeqi Luo$^1$, Lipeng Ma$^1$, \\ \textbf{Ben Fei$^{3, }$\Envelope}, \textbf{Lei Bai$^{2, }$\Envelope}\\
	$^1$Fudan University, $^2$Shanghai AI Laboratory, $^2$Chinese University of Hong Kong\\
    \texttt{jyxu22@m.fudan.edu.cn}, 
    \texttt{wdyang@fudan.edu.cn},
    \texttt{benfei@cuhk.edu.hk}, 
    \texttt{baisanshi@gmail.com}\\
    $^{*}$ Equal Contributions, \Envelope Corresponding Authors
}

\date{}
\renewcommand{\shorttitle}{IceDiff}

\hypersetup{
pdftitle={A template for the arxiv style},
pdfsubject={q-bio.NC, q-bio.QM},
pdfauthor={David S.~Hippocampus, Elias D.~Striatum},
pdfkeywords={First keyword, Second keyword, More},
}

\begin{document}
\maketitle

\begin{abstract}
Variation of Arctic sea ice has significant impacts on polar ecosystems, transporting routes, coastal communities, and global climate. 
Tracing the change of sea ice at a finer scale is paramount for both operational applications and scientific studies. 
Recent pan-Arctic sea ice forecasting methods that leverage advances in artificial intelligence has made promising progress over numerical models. However, forecasting sea ice at higher resolutions is still under-explored.
To bridge the gap, we propose a two-staged deep learning framework, IceDiff, to forecast sea ice concentration at finer scales. 
IceDiff first leverages an independently trained vision transformer to generate coarse yet superior forecasting over previous methods at a regular 25km x 25km grid.
This high-quality sea ice forecasting can be utilized as reliable guidance for the next stage.
Subsequently, an unconditional diffusion model pre-trained on sea ice concentration maps is utilized for sampling down-scaled sea ice forecasting via a zero-shot guided sampling strategy and a patch-based method. 
For the first time, IceDiff demonstrates sea ice forecasting with the 6.25km x 6.25km resolution.
IceDiff extends the boundary of existing sea ice forecasting models and more importantly, its capability to generate high-resolution sea ice concentration data is vital for pragmatic usages and research.
\end{abstract}

\section{Introduction}
\label{sec:intro}

Arctic sea ice is essential for global climate change and local communities. Although the exact teleconnections between mid-latitude weather and the variation of Arctic sea ice are yet to be revealed, the reduction of sea ice during the last few decades clearly has impacts that transcend the Arctic region ~\citep{wang2019subseasonal,andersson2021seasonal}. Hence, establishing skillful forecasting models is paramount for both transportation in polar regions and scientific research in geoscience. 

Existing physics-based numerical models have demonstrated the ability to accurately forecast sea ice concentration (SIC, value ranges from 0\% to 100\%) for short lead times of a few weeks.
However, when forecasting at longer scales, their performance significantly decreases, and their forecasting skills are even worse than those of simple statistical models~\citep{andersson2021seasonal}. 
With the advent of deep learning methods and their showcased capability of handling complex visual and natural language data at large scale, a series of works based on deep learning approaches have been committed to SIC forecasting with promising results. 
Notable studies cover forecasting leads range from daily to seasonal scale, e.g., IceNet~\citep{andersson2021seasonal} provides accurate sea ice extent (SIE, areas where the SIC value is greater than 15\% and could be covered with sea ice) forecasts for 6 months lead time. SICNet~\citep{ren2022data} and SICNet$_{90}$~\citep{ren2023predicting} are skillful for forecasting SIC in subsequent 7 and 90 days, respectively. 
Despite the improved forecasting skills of deep learning models over previous approaches, forecasting sea ice at finer scales, which is crucial, especially for operational applications, is still under-explored. 
SwinRDM~\citep{chen2023swinrdm} with similar intention has been proposed for weather forecasting. 
It combines an improved SwinRNN~\citep{hu2023swinvrnn} model with a denoising diffusion probabilistic model (DDPM)~\citep{ho2020denoising} to achieve super-resolution (SR) in forecasting climate variables.

In this study, we propose IceDiff, a two-staged deep learning framework, that generates skillful forecasts at three different temporal scales with spatially down-scaled high-quality samples. In the first stage, it leverages U-Net-based vision transformer and provides forecasts that are superior to existing deep learning models. 
Then, the down-scaling task is implemented using well-trained DDPM as an effective prior and guided by forecasts at the original scale. 
Our DDPM is pre-trained on daily SIC maps with a resolution of 256 x 256, and it can also generalize well to weekly and monthly SIC maps.
In every step $t$ of the sampling process of DDPM, the intermediate variable ${x}_0$ is estimated from the denoised SIC map ${x}_t$.
Then we propose a convolution kernel with optimizable parameters to simulate the up-scale degradation for the intermediate variable ${x}_0$, which results in up-scaled $\tilde{x}_0$.
A distance function between the low-resolution SIC map and up-scaled $\tilde{x}_0$ is utilized to dynamically update the parameters of the convolution kernel in the reverse steps and guide the sampling of SIC map in next time step.
Furthermore, to make our IceDiff suitable for SIC maps with various resolutions, a patch-based method is also designed for down-scaling any size SIC maps. 
We show in this work that using low-resolution SIC maps as guidance can better capture small-scale details in the generated high-resolution SIC maps.

Our contributions are three folds:
\begin{itemize}
    \item 
    We propose a novel SIC forecasting framework IceDiff that demonstrates skillful in three different temporal scales which cover subseasonal to seasonal forecasts, specifically, 7 days, weekly average of 8 weeks and monthly average of 6 months, that are commonly investigated by the community. 
    IceDiff achieves state-of-the-art results in all three scales than existing deep learning models.  
    
    \item
    IceDiff devises an optimizable convolutional kernel to simulate the up-scaling and designs a patch-based method for down-scaling any-size SIC maps with a single and fixed-resolution unconditional diffusion model.
    By utilizing the powerful generative diffusion prior inherent in this diffusion model, the high-resolution SIC maps can be obtained through zero-shot sampling. 
    Moreover, our IceDiff is capable of capturing small-scale details and generating high-resolution SIC maps with wide applicability.

    \item 
    Extensive experiments show that our IceDiff could provide accurate SIC forecasts and generate high-quality sea ice samples with finer details. IceDiff not only provides a new and comprehensive baseline for data-driven sea ice research but also extends the capability of existing SIC forecasting models and pushes this area of research towards more operational usage.
    
\end{itemize}

\section{Related Works}
\label{sec:related}

\subsection{Deep Learning-based Sea Ice Forecasting}
Deep learning models have drawn the attention of sea ice research communities and been widely investigated for Arctic sea ice forecasting~\citep {petrou2019prediction,kim2020prediction,ali2021sea,ali2022mt}.
As a pioneering work, ConvLSTM ~\citep{liu2021extended} predicts SIC within the region of Barents Sea and has showcased comparable results to the state-of-the-art numerical climate models.
IceNet~\citep{andersson2021seasonal} employs the U-Net architecture for merging pre-selected ERA5 variables and SIC data to improve the accuracy of seasonal SIE prediction up to six months. The experimental results have shown that the forecasting skill of IceNet in pan-Arctic region is superior to the numerical sea ice simulation model SEAS5. MT-IceNet~\citep{ali2022mt} adopts similar approach of IceNet and adds bi-monthly data to further improve forecasting SIC at seasonal scale.
Mu et al.~\citep{mu2023icetft} propose to merge atmospheric and oceanographic in a weighted manner through a variable selection network and provide nine months of forecasting leads. 
Since SIC can be derived from SIC, SICNet~\citep{ren2022data} is proposed to solely rely on SIC data to predict concentration in the next seven days. 
Similar to IceNet, SICNet is built upon U-Net architecture and added with channel-wise dependency via temporal-spatial attention module (TSAM), which is essentially a revised version of the convolutional block attention module.
By fusing the output of two SICNet, i.e. one for modeling SIC and the other for extracting climate information, and adjusting the number of output channels, Ren et al.~\citep{ren2023predicting} extends the lead time to 90 days.

Besides CNN and LSTM-based models, recent works leverage the latest advances in artificial intelligence to further improve forecasting performance. 
For instance, IceFormer \citep{zheng2024spatio} jointly models SIC and sea ice thickness (SIT) based on Transformer backbone. It first decomposes high dimensional spatio-temporal data into sequences via multivariate empirical orthogonal functions~\citep{sparnocchia2003multivariate,shao2021ocean}. Then an encoder-decoder Transformer is utilized to predict SIC and SIT for up to 45 days.

Despite the performance gain of aforementioned deep learning models, non of them have addressed forecasting sea ice at scales lower than 25km.


\subsection{Super Resolution}


Super-resolution aims to reconstruct high-resolution (HR) images from low-resolution (LR) counterparts. 
For this purpose, several works~\citep{liu2020unsupervised,liu2020photo} explore the use of Variational Autoencoders for perceptual image super-resolution.
These methods usually improve the restoration quality of image super-resolution by learning the conditional distribution of high-resolution images induced by low-resolution images. 
Although these methods are able to achieve realistic perceptual quality, VAE-based methods are often not as satisfactory due to their limited generation capabilities. 
To this end, other deep learning methods, such as deep convolutional neural networks, have been widely used for super-resolution tasks in various domains. 
Convolutional neural networks (CNNs) based methods typically use HR images corresponding to LR images to supervise training, which utilizes reconstruction loss to train CNNs. 
\cite{zhou2019kernel,anwar2020deep} propose and optimize their CNN-based methods to tackle various super-resolution tasks, improving their generalization ability and robustness on real photos.
However, the network trained solely on reconstruction loss makes it difficult to capture detailed information in images and generate diverse image results. 
Therefore, Generative Adversarial Networks (GANs) are proposed to solve these challenges.
Leinonen et al~\citep{leinonen2020stochastic} introduced a recurrent, stochastic super-resolution GAN for down-scaling time-evolving atmospheric fields, demonstrating the potential of GANs in improving the spatial resolution of LR images in atmospheric sciences applications. 
These models encourage the generator to generate high-quality images that the discriminator cannot distinguish from real images. 
Although GAN-based methods can attain satisfactory down-scaled results, the diversity of these methods could be enhanced. 

Currently, diffusion model-based methods have been more widely used due to their diversity in generating images and the ability to generate high-quality images comparable to GANs~\citep{tu2024taming,fei2023generative}. 
For instance, Chen et al.~\citep{chen2023swinrdm} utilize super-resolution based on the conditional model to recover the high spatial resolution and finer-scale atmospheric details, unfolding the ability of diffusion models to generate high-quality and detailed images in super-resolution tasks.

The objective of this study is to forecast SIC maps at a high resolution of 6.25 km. 
To achieve this, we have improved the SwinTransformer for predicting SIC at the standard resolution comparable to current methods.
Additionally, we employ a diffusion-based super-resolution model to generate high-resolution and high-quality results via zero-shot sampling.

\begin{figure*}[t]
    \centering
    \includegraphics[width=\linewidth]{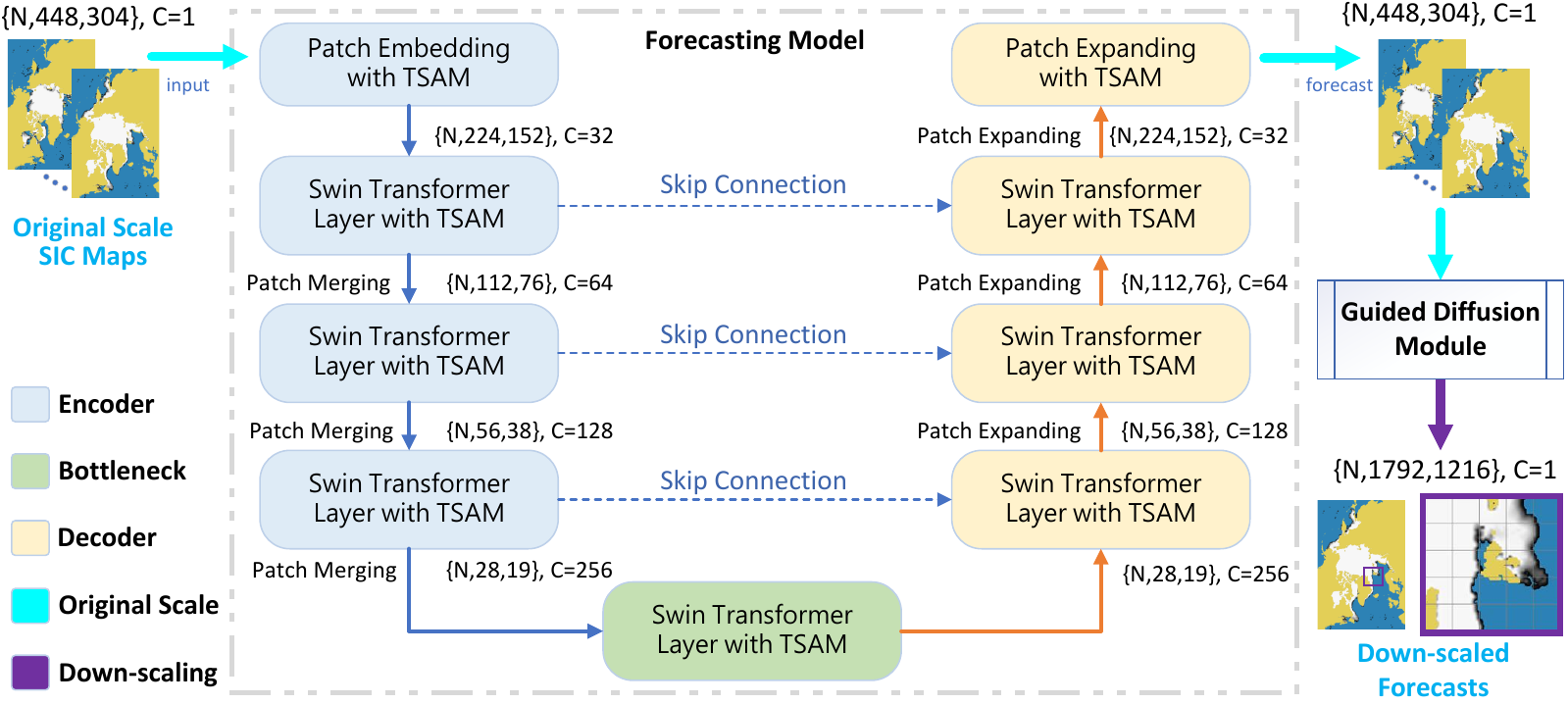}
    \caption{\textbf{Overview of IceDiff Framework.} IceDiff first employs the forecasting model (FM) that takes history SIC data as input to generate subsequent predictions at 25km x 25km grid. N denotes the number of sequential time stamps, for instance, set N to 7 for seven days of forecasting leads, FM takes a tensor with shape of $\{7, 448, 304 \}$ as input and forecasts the SIC for the next 7 days. C represents the quantity of embedded channels within each layer of FM, the original SIC map data has a channel of 1 rather than 3 for regular images. The guided diffusion module (GDM) then leverages accurate forecasts as guidance to generate downscaled samples at 6.25km x 6.25km grid with high quality, see details in Figure~\ref{fig:enter-label}.}
    \label{fig:framework}
\end{figure*}

\section{Methodology}
\label{sec:methods}
Our IceDiff framework comprises two components, i.e. the forecasting model (FM) and the guided diffusion module (GDM), as depicted in Figure~\ref{fig:framework}.
We elaborate on design details in the following sections.

\begin{figure*}[t]
    \centering
    \includegraphics[width=\linewidth]{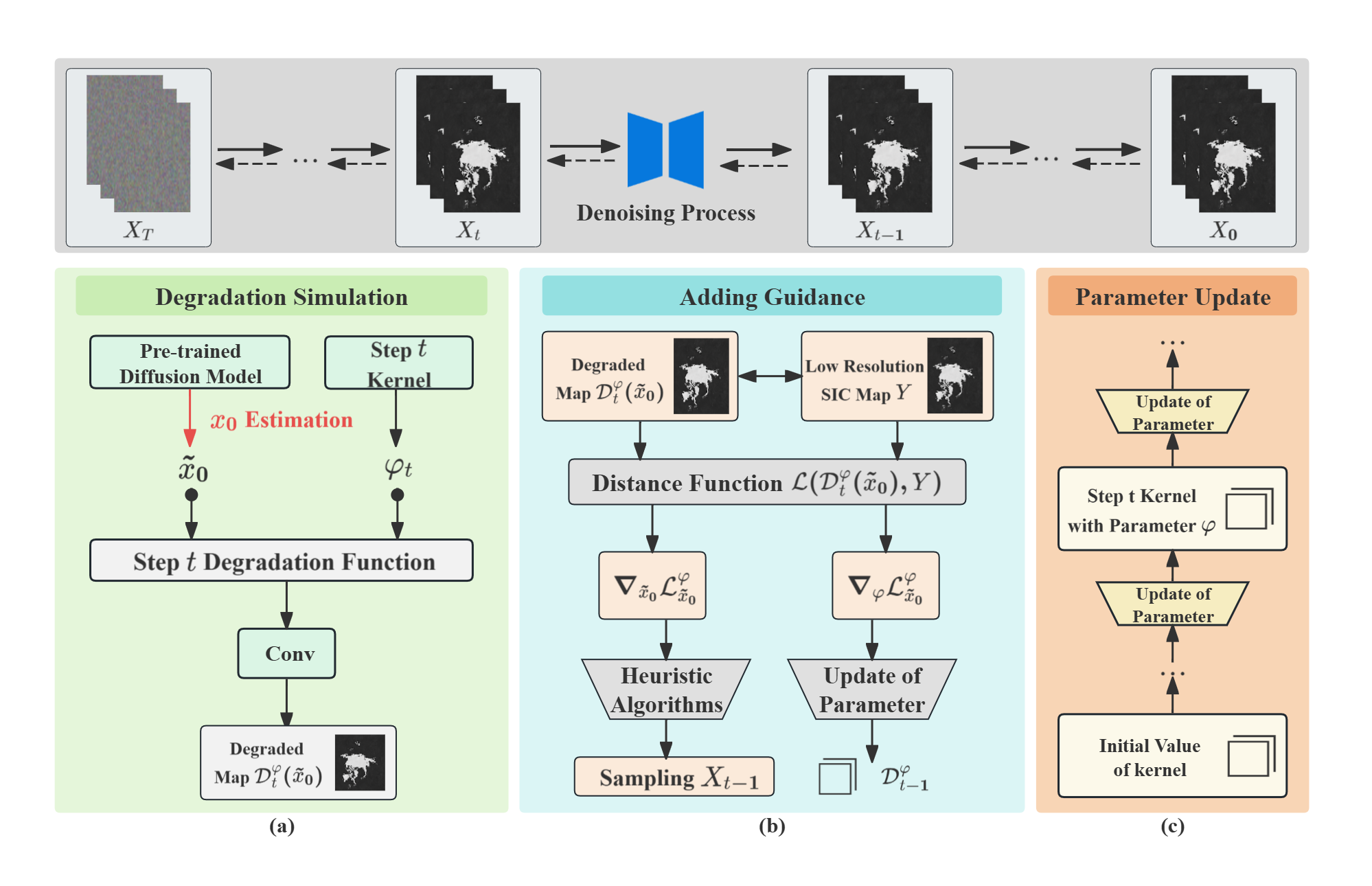}
    \caption{\textbf{Overview of GDM for SIC map dowdnscaling.} \textbf{(a)} Pre-trained diffusion model is used to estimate $\tilde{x}_0$ at reverse step $t$. The optimizable convolutional kernel of step $t$ reduces the resolution of $X_t$ so that guidance from low-resolution SIC maps can be incorporated to sample $X_{t-1}$.
    \textbf{(b)} Low-resolution SIC map $Y$ is integrated during the sampling process. A distance function is introduced to describe the distance between a low-resolution SIC map and the generated high-resolution SIC map $x$ after convolution. The gradient of distance function could be used to guide the sampling of $x_{t-1}$ and \textbf{(c)} dynamically update the parameter of the convolutional kernel.}
    \label{fig:enter-label}
\end{figure*}
                                                                                                                                                                                                                                                                                                                                                                                                                                                                                                                                                                    
\subsection{Forecasting Model}
Swin Transformer~\citep{liu2021swin} as a vision transformer backbone has been successfully adopted by various data-driven weather forecasting methods~\citep{chen2023swinrdm,zhang2014fuxi,bi2023accurate}. Inspired by their efforts and the verified effectiveness of TSAM~\citep{ren2022data}, we propose to construct FM by utilizing the U-Net architecture~\citep{cao2022swin} and building on top of the Swin Transformer V2 backbone~\citep{liu2022swin} with ResNet$_{TSAM}$ block~\citep{ren2022data} for better capturing both spatial and temporal features of SIC.
As illustrated in Figure.~\ref{fig:framework}, the FM consists of \textit{encoder}, \textit{bottleneck} and \textit{decoder}. For the patch embedding layer, it first applies a 2 x 2 windows size to partition the original SIC input into patches and maps them into tokens via 2D convolution. Then, the temporal and spatial dependencies are modeled through the ResNet$_{TSAM}$ block. 
Both the encoder and the decoder has 3 layers of Swin Transformers that comprises of pre-configured even number of Swin Transformer Blocks for feature extraction and restoration. The calculation of two consecutive blocks can be described as follows:
\begin{align}
	&z^{b}_{s} = LN(WMSA(z^{b-1})) + z^{b-1}, \nonumber \\
	&z^{b} = LN(MLP(z^{b}_{s})) + z^{b}_{s}, \nonumber \\
	&z^{b+1}_{s} = LN(SWMSA(z^{b})) + z^{b}, \nonumber \\
	&z^{b+1} = LN(MLP(z^{b+1}_{s})) + z^{b}_{s},
\end{align}
where $z^{b}_{s}$ and $z^{b}$ represents the output feature of the (\underline{\textbf{S}}hifted) \underline{\textbf{W}}indow-\underline{\textbf{M}}ulti-head \underline{\textbf{S}}elf \underline{\textbf{A}}ttention module and the MLP module for block $b$, respectively~\citep{liu2022swin}; 
LN denotes the LayerNorm operation.
The combination of such two consecutive blocks, i.e. the former for the employment of WMSA and the latter for SWMSA, ensures that local patterns of adjacent grids are sufficiently attended by the backbone network.
We further adopt a shift window size of 28 x 19 that is consistent with the ratio of SIC input map for facilitating the SWMHA mechanism. The output of the last block within each layer is delivered to a ResNet$_{TSAM}$ block for establishing channel-wise dependencies. Before the subsequent encoder and decoder layers, patch merging and patch expanding~\citep{cao2022swin} are respectively performed to reshape the feature and fit it to the desired size. Following prior works~\citep{cao2022swin,ren2022data,andersson2021seasonal}, the skip connection is added between the corresponding encoder and decoder layers to improve the performance. The final prediction is generated via a patch expanding layer with ResNet$_{TSAM}$ that up-samples the decoded feature to the size of the original SIC input.

\subsection{Exploiting Generative Diffusion Prior for High-resolution SIC Maps via Zero-shot Sampling}

\begin{figure*}
    \centering
    \includegraphics[width=\linewidth]{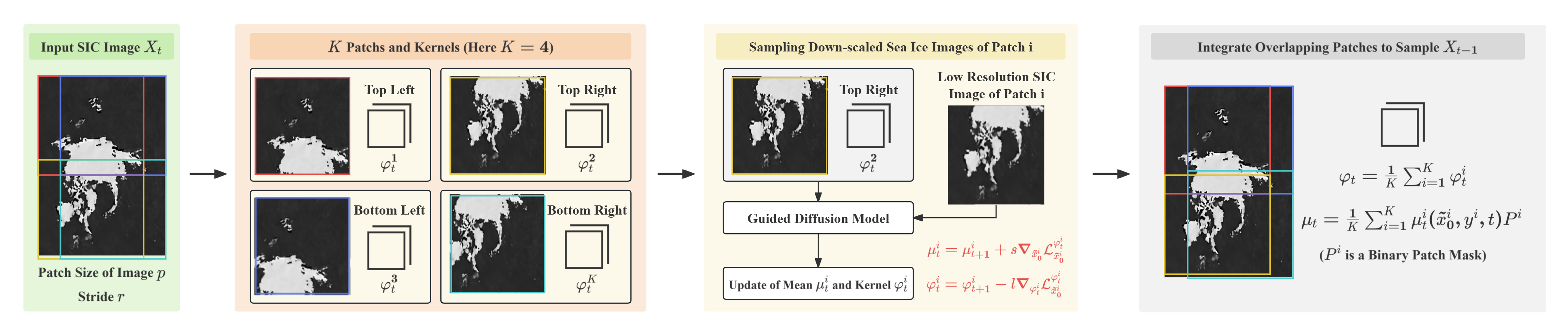}
    \caption{\textbf{Illustration of the patch-based diffusion model pipeline.} We set patch size $p=256$ and stride $r=128$ to fit the size of pre-trained models. This setting will divide the example map into $K=4$ sub maps boxed with different colors. For each sub map, individual optimizable convolution kernel with adaptive parameters is set to add guidance from corresponding low resolution SIC sub map. At each sampling step $t$, the average estimated noise on four overlapping blocks will be utilized to update the mean $\mu$ used to sample $x_{t-1}$.}
    \label{fig:patch-based}
\end{figure*}

In this section, we exploit a down-scaling component based on an unconditional diffusion model pre-trained on SIC maps to generate realistic high-resolution SIC maps with rich details. 
Off-the-shelf methods are limited to the forecasting of SIC maps at a scale of 25km.
However, small-scale SIC maps remain unexplored which is of significant value to ecosystems, transporting routes, coastal communities and global climate.
Our IceDiff-GDM down-scales the SIC forecasts by IceDiff-FM to a scale of 6.25km, further expanding the accuracy and practicality of sea ice prediction. 

Specifically, the reverse process of the diffusion model is conditioned on the low-resolution sea ice map $y$, which transforms distribution $p_\theta(x_{t-1}|x_t)$ into conditional distribution $p_\theta(x_{t-1}|x_t, y)$. 
Previous work \citep{dhariwal2021diffusion} have derived the conditional transformation formula in the reverse process:
\begin{align}
\log_{}&{p_\theta(x_{t}|x_{t+1},y)}=\log_{}{(p_\theta(x_{t}|x_{t+1})p(y|x_{t})) }+N_1 \\
&\approx \log_{}{p_\theta(z)+N_2}\;\;  z\sim \mathcal{N}(z;\mu_\theta(x_t,t)+\Sigma g,\Sigma I),
\end{align}
where $g$ refers to $\nabla_{x_t}\log_{}{p(y|x_t)}$. ${N_1=\log{}{p_\theta(y|x_{t+1})}}, {N_2}$ is a constant related to $g$ (proof in Appendix). The variance of the reverse process $\Sigma= \Sigma_\theta (x_t)$ is constant. 

Therefore, sampling process of $p_\theta(x_{t-1}|x_t,y)$ integrates the gradient term $g$ with the diffusion model $(\mu,\Sigma)=(\mu_\theta(x_t),\Sigma_\theta(x_t))$, which plays a role in controlling the direction of map generation. 
We use a heuristic algorithm to approximate the value of $g$:
\begin{align}
&\log_{}{ p(y\mid x_{t})}=- \log_{}{N}-s\mathcal{L}( \mathcal{D}(\tilde{x}_0),y)\\
g&=\nabla _{x_t}\log_{}{p(y|x_t)}=-s\nabla_{x_t}\mathcal{L}(\mathcal{D}(x_t),y).
\end{align}

Among them, $s$ is the scaling factor used to control the degree of guidance, and $K$ is the normalization factor, which is the constant $p_\theta(y|x_{t+1})$. 
And $\mathcal{L}$ is a distance function used to measure the distance deviation between two maps. 
$\mathcal{D}$ is an up-scale convolution function with optimizable parameters, where the parameters dynamically optimized with the gradient of distance function at every reverse step. 
$\mathcal{D}$ connects two different resolution SIC maps, allowing low-resolution SIC maps to provide more effective guidance. 

\begin{algorithm}[t]\small
\renewcommand{\algorithmicrequire}{\textbf{Input:}}
\renewcommand{\algorithmicensure}{\textbf{Output:}}
\caption{Guided diffusion model with the guidance of low-resolution SIC map $y$. An unconditional diffusion model pre-trained on SIC maps $\epsilon_\theta(x_t,t)$ is given.}
\label{alg.1}
\begin{algorithmic}[1]
\REQUIRE Low-resolution SIC map $y$, downscale convolution function $\mathcal{D}$ composed of optimized convolutional kernels $\mathcal{K}$ with parameters $\varphi$, learning rate $l$, guidance scale $s$, distant function $\mathcal{L}$.
\ENSURE Output high-resolution SIC map $x_0$ conditioned on $y$.
Sample $x_T$ from $\mathcal{N}(0,I)$

    \FORALL{t from T to 1}
        \STATE $\tilde{x} _0=\frac{x_t}{\sqrt{\bar{\alpha}_t }}-\frac{\sqrt{1-\bar{\alpha}_t}\epsilon_\theta(x_t,t)}{\sqrt{\bar{\alpha}_t }}$\
        
        \STATE $\mathcal{L}_{\varphi,\tilde{x}_0} =\mathcal{L}(y,\mathcal{D}^{\varphi}(\tilde{x}_0))$\

        \STATE $\tilde{x}_0 \gets \tilde{x}_0-\frac{s(1-\bar{\alpha}_t) }{\sqrt{\bar{\alpha}_{t-1}}\beta_t}\nabla_{{\tilde{x}}_0}\mathcal{L}_{\varphi,\tilde{x}_0}$\
        \STATE $\tilde{\mu}_t=\frac{\sqrt{\bar{\alpha}_{t-1}}\beta_t}{1-\bar{\alpha}_t}\tilde{x}_0+\frac{\sqrt{\bar{\alpha}_{t}}(1-\bar{\alpha}_{t-1})}{1-\bar{\alpha}_t}{x}_t$\

        \STATE $\tilde{\beta}_t=\frac{1-\bar{\alpha}_{t-1}}{1-\bar{\alpha}_t}\beta_t$\
        
        \STATE Sample $x_{t-1}$ from $\mathcal{N}(\tilde{\mu}_t,\tilde{\beta}_tI)$\

        \STATE $\varphi \gets \varphi-l\nabla_{\varphi}\mathcal{L}_{\varphi,\phi,\tilde{x}_0}$\
    \ENDFOR
\STATE \textbf{return}  $x_0$
    \end{algorithmic}
\end{algorithm}

As shown in \Cref{alg.1}, GDM undergoes $T$ reverse steps to gradually restore pure Gaussian noise $x_T\sim \mathcal{N}(0, I)$ to high-resolution SIC maps. 
In each reverse step $t$, the diffusion model calculates the instantaneous estimated value $\tilde{x}_0$, which then undergoes the up-scale convolution function $\mathcal{D}$ at time $t$ to establish a distance metric $\mathcal{L}$ with the low-resolution SIC map. 
The gradients of distance function $\mathcal{L}$ with respect to $\tilde{x}_0$ and the convolution kernel parameters of the up-scale convolution function are used to update $\tilde{x}_0$ and the convolution kernel parameters, respectively. 
The update of convolution kernel parameters ensures the accuracy of the model's up-scaling process, making guidance of low-resolution SIC maps more precise and effective. 

Furthermore, pre-trained models can only generate fixed-size SIC maps. 
To tackle the problem of sampling down-scaled SIC forecasts at any resolution, we employ patch-based methods (Figure~\ref{fig:patch-based}). 
By leveraging this strategy, our model is capable of down-scaling sea ice maps to desired size
and the practicality of the model has been further enhanced.
More details about our patch-based strategy could be found in Appendix.

\section{Experiments}
\label{sec:exp}

\begin{table*}[t]\small
  \centering
  \caption{\textbf{Quantitative results of forecasts}. IceDiff-FM is evaluated on three different temporal scales. The output length of IceDiff-FM equals a pre-defined forecasting lead as depicted in Figure~\ref{fig:framework}. We re-implement SICNet in Pytorch and train with the same dataset to compare with our IceDiff-FM. Other deep learning methods marked with * represent that we report their experiment results from the original paper due to code and dataset accessibility reasons, e.g. the source code for IceFormer and SICNet$_{90}$ are not available. Note that the performance of IceNet is reported from the MT-IceNet paper. Despite the potential differences that may caused by the source of the dataset, IceDiff-FM still exhibits comparable and superior results in all three temporal scales. The validated accuracy of IceDiff-FM could provide reliable guidance for IceDiff-GDM to generate high-quality down-scaled SIC.}
  \label{tab:main}
\begin{tabular}{c|l|l|cccccc}
\toprule[1pt]
Temporal Scale               & \multicolumn{1}{c|}{Lead Time}           & \multicolumn{1}{c|}{Methods}&RMSE$\downarrow$&MAE$\downarrow$& $R^{2}$$\uparrow$&NSE$\uparrow$&IIEE$\downarrow$&SIE$_{dif}$$\downarrow$\\ \hline
\multirow{5}{*}{Subseasonal} & \multirow{2}{*}{7 Days (Daily)}          & SICNet           & 0.0490 & 0.0100 & 0.982 & 0.979 & 976 & 0.0718\\
                             &                                          & IceDiff-FM       &\textbf{0.0396} & \textbf{0.0080} & \textbf{0.989} & \textbf{0.987} & \textbf{835} &  \textbf{0.0315}\\ \cline{2-9} 
                             & 45 Days (Daily)                          & IceFormer*       &0.0660 & 0.0201 & 0.960 & - & - &  -   \\ \cline{2-9} 
                             & 8 Weeks Average (Weekly)                    & IceDiff-FM       &\textbf{0.0553}& \textbf{0.0112}&\textbf{0.973}&\textbf{0.969}&\textbf{1353}&\textbf{0.0871}\\ \cline{2-9} 
                             & 90 Days (Daily)                          & SICNet$_{90}$*   &   -   & 0.0512 & - & - & - &  -   \\ \hline
\multirow{3}{*}{Seasonal}    & \multirow{3}{*}{6 Months Average (Monthly)} & IceNet*          &   0.1820   & 0.0916 & 0.567 & - & - &   -  \\
                             &                                          & MT-IceNet*       &   0.0777   & 0.0197 & 0.915 & - & - &   -  \\
                             &                                          & IceDiff-FM       &\textbf{0.0648}&\textbf{0.0168}&\textbf{0.919}&\textbf{0.913}&\textbf{2016}&\textbf{0.2657}\\ 
\bottomrule[1pt] 
\end{tabular}
\end{table*}

%
%
%


\begin{table}[t]\small
\centering
\caption{\textbf{Quantitative comparison between IceDiff-GDM for SIC down-scaling and other methods}. Const. and FID represents the Consistency and Fréchet Inception Distance metrics, respectively. 7.D., 8.W.A. and 6.M.A. are abbreviations for 7 days, 8 weeks average and 6 months average for different forecasting leads as in Table~\ref{tab:main}.}
\resizebox{0.5\textwidth}{!}{
\begin{tabular}{c|c c|c c| c c}
    \toprule[1pt]
     \multirow{2}{*}{\textbf{Method}} &\multicolumn{2}{c|}{7.D.} & \multicolumn{2}{c|}{8.W.A.}&\multicolumn{2}{c}{6.M.A.}\\
     \cmidrule(lr){2-7}

    &Const.$\downarrow$&FID$\downarrow$
    &Const.$\downarrow$&FID$\downarrow$
    &Const.$\downarrow$&FID$\downarrow$\\
    \midrule
    Nearest&20.01&78.45&21.98&80.17&23.07&88.36 \\
    Bilinear&17.88&70.75&20.02&77.29&18.31&80.73 \\
    BiCubic&13.13&54.71&13.40&68.85&10.09&70.18 \\
    DGP&23.18&75.42&28.62&81.42&29.74&85.47 \\
    GDP&8.26&40.15&10.54&52.40&10.96&55.72 \\
    \midrule 
    IceDiff-GDM&\textbf{6.84}&\textbf{29.34}&\textbf{7.25}&\textbf{31.29}&\textbf{6.62}&\textbf{39.79} \\
    \bottomrule[1pt]

  \end{tabular}
  }
 \label{GDM_metrics}
\end{table}

\begin{figure}[t]
    \centering
    \includegraphics[width=\linewidth]{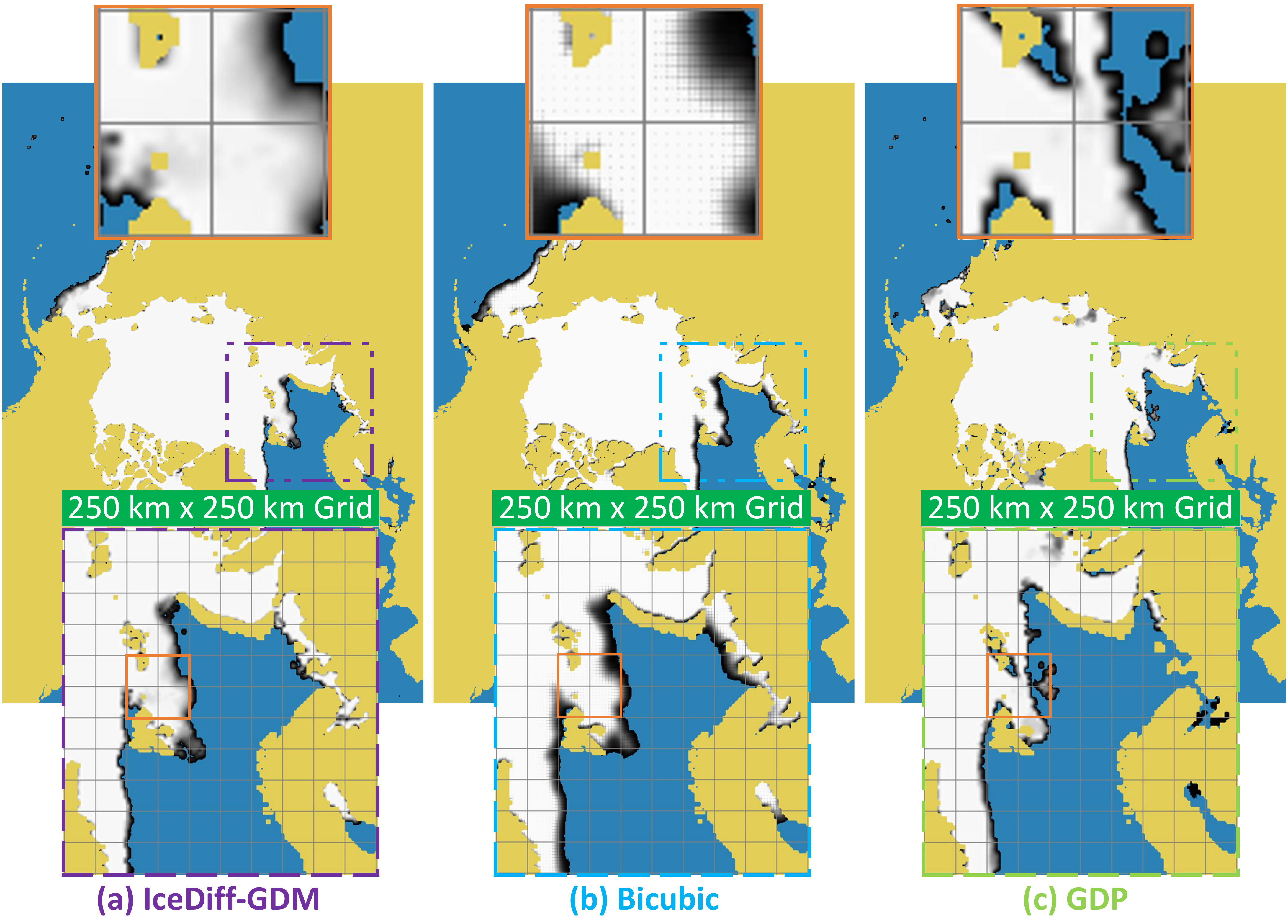}
    \caption{\textbf{Qualitative analysis of down-scaled SIC map.} We visualize the SIC map by coloring the land and the ocean in yellow and blue, respectively. Since SIC has a range of $[0,1]$, we map it to $[0,255]$ and replicate it for all 3 RGB channels. The darker the pixels appear, the SIC values are more closer to 0\%. For better identification of the difference, we crop out a region and plot a grid size of 250km and further displays details on 4 x 4 grid.}
    \label{fig:high_res}
\end{figure}

\subsection{Dataset and Baseline Setup}
\label{exp:data}
We evaluate our proposed IceDiff framework on the G02202 Version 4 dataset from the National Snow and Ice Data Center (NSIDC). It records daily SIC data starting from October 25$^{th}$ 1978 and provides the coverage of the pan-Arctic region (N:-39.36$^{\circ}$, S:-89.84$^{\circ}$, E:180$^{\circ}$, W:-180$^{\circ}$). 
The SIC map is formed of 448 x 304 pixels and each of which represents an area of 25km x 25km grid and has a value ranging from 0\% to 100\%. 
We choose data from October 25$^{th}$ 1978 to 2013 as the training dataset, the years 2014 and 2015 are selected as validation set, and data from 2016 to 2023 are used for testing. 
IceDiff-FM is trained on three different temporal scales, i.e. the input of SIC maps are 7 days, 8 weeks average and 6 months average, covering forecasting leads from subseasonal to seasonal scale. Corresponding deep learning-based models are selected for comparison, as illustrated in Table~\ref{tab:main}. Note that we re-implemented SICNet (the source code is not available) based on the original paper and trained under an environment identical to our FM.
To evaluate our IceDiff-GDM, five methods are selected to set up baselines for down-scaling quality comparison. 

\subsection{Implementation Details}
\label{sec:implem}
We independently trained IceDiff-FM on SIC dataset of three aforementioned temporal scales for 60 epochs. For each temporal scale, IceDiff-FM outputs the same length of input data as forecasting leads, i.e. 7 days, 8 weeks average, and 6 months averages. The dimensions of embedding features of encoder, decoder, and bottleneck are all set to 32. We utilize the same hyperparameter of TSAM as in~\citep{ren2022data}.
IceDiff-FM is optimized by AdamW using Pytorch on one NVIDIA A100 80GB GPU for all experiments with a learning rate of 0.0005. 
All daily SIC maps are resized and cropped to 256 x 256 to train our IceDiff-GDM.
We train our IceDiff-GDM using AdamW with $\beta_1=0.9$ and $\beta_2=0.999$. We train in 16-bit precision using loss-scaling, but maintain 32-bit weights, EMA, and optimizer state. We use an EMA rate of 0.9999. When sampling with 1000 timesteps, we use the same noise schedule as for training.

\begin{table}[t]\small
  \centering
  \caption{\textbf{Ablation study on TSAM.} We compare the forecasting results between IceDiff-FMs with or without the TSAM. L.T. represents the forecasting leads as in Table~\ref{tab:main}. We utilize the same abbreviations in Table~\ref{tab:abla.io}.}
  \label{tab:abla.tsam}
  \resizebox{0.5\textwidth}{!}{
  \begin{tabular}{c|c|c c c c c c}
    \toprule[1pt]
L.T. & TSAM
&RMSE$\downarrow$&MAE$\downarrow$&$R^{2}$$\uparrow$&NSE$\uparrow$&IIEE$\downarrow$&SIE$_{dif}$$\downarrow$\\ 
    \hline
  \multirow{2}{*}{7.D.} 
  &\XSolid& 0.0403 & 0.0082 & 0.9881 & 0.9860 & 874 & 0.0336\\ 
  &\Checkmark& \textbf{0.0396} & \textbf{0.0080} & \textbf{0.9885} & \textbf{0.9865} & \textbf{835} &  \textbf{0.0315} \\ 
    \hline

  \multirow{2}{*}{8.W.A.}
  &\XSolid&\textbf{0.0535}&0.0116&\textbf{0.975}&\textbf{0.971}&1386&0.1420\\
  &\Checkmark&0.0553&\textbf{0.0112}&0.973&0.969&\textbf{1353}&\textbf{0.0871}\\
    \hline

  \multirow{2}{*}{6.M.A.}
  &\XSolid&0.0763&0.0194&0.883&0.875&2610&0.4919\\
  &\Checkmark&\textbf{0.0648}&\textbf{0.0168}&\textbf{0.919}&\textbf{0.913}&\textbf{2016}&\textbf{0.2657}\\
    \bottomrule[1pt]
  \end{tabular}
}
\end{table}

\subsection{Evaluation Metrics}
\label{sec:metric}
To evaluate IceDiff-FM, we select commonly used root mean square error (RMSE) and mean absolute error (MAE) for comparison of forecasting accuracy. We also leverage $R^{2}$ score to evaluate the performance:
\begin{align}
R^{2} = 1 - \frac{RSS}{TSS}.
\end{align}
where RSS represents the sum of squares of residuals and TSS denotes the total sum of squares. The Integrated Ice-Edge Error score~\citep{goessling2016predictability} is introduced to evaluate the SIE (where the SIC value is greater than 15\%) prediction:
\begin{align}
IIEE & = O + U, \\
O & = SUM(Max(SIE_{p} - SIE_{t},0)),\\
U & = SUM(Max(SIE_{t} - SIE_{p},0)),
\end{align}
\begin{equation}
S I E_p, S I E_t=\left\{\begin{array}{l} 
{1, S I C>15}
\\
0, S I C \leq 15
\end{array}{ }\right.
\end{equation}
where O and U represent the overestimated and underestimated SIE  between the prediction ($SIE_{p}$) and the ground truth ($SIE_{t}$), respectively. The difference between the forecasted and ground truth sea ice area (in millions of $km^2$) is calculated as follows:
\begin{align}
    SIE_{dif} = \frac{SUM(|SIE_{p}-SIE_{t} |) \times 25 \times 25}{1000000}.
\end{align}

We also adopt the Nash-Sutcliffe Efficiency~\citep{NASH1970282} to further evaluate the predicted quality:
\begin{align}
    NSE = \frac{1-SUM((SIC_{t} - SIC_{p})^2)}{SUM( (SIC_{t} - Mean(SIC_t))^2 )}
\end{align}

As to the evaluation of down-scaling quality, we utilize the Fréchet Inception Distance (FID) \citep{NIPS2017_8a1d6947} and Consistency for comparison.
In addition, we adopt Consistency to quantify the faithfulness between the generated map and the low-resolution map.

\begin{table}[t]\small
  \centering
  \caption{\textbf{Ablation study on the ratio of Input/Forecast number of SIC maps}. Hf. Db. and Eq. represent that the FM takes input data which has \textit{half}, \textit{double}, and \textit{equal} length of forecasting leads, respectively. For the Hf. of 7 Days, we set the input length to 3. }
  \label{tab:abla.io}
  \resizebox{0.5\textwidth}{!}{
  \begin{tabular}{c|c| c c c c c c}
    \toprule[1pt]
L.T. & I./F.
&RMSE$\downarrow$&MAE$\downarrow$&$R^{2}$$\uparrow$&NSE$\uparrow$&IIEE$\downarrow$&SIE$_{dif}$$\downarrow$\\ 
    \hline
  \multirow{3}{*}{7.D.} 
  &Hf.&0.0416&0.0084&0.987&0.985&905&0.0538\\ 
  &Db.&0.0401&0.0084&0.988&0.986&889&0.0599\\ 
  &Eq.&\textbf{0.0396}&\textbf{0.0080}&\textbf{0.989}&\textbf{0.987}&\textbf{835}&\textbf{0.0315}\\ 
    \hline

  \multirow{3}{*}{8.W.A.}
  &Hf.&0.0580&0.0118&0.970&0.966&1405&0.1176\\
  &Db.&0.0594&0.0121&0.969&0.964&1435&0.1221\\
  &Eq.&\textbf{0.0553}&\textbf{0.0112}&\textbf{0.973}&\textbf{0.969}&\textbf{1353}&\textbf{0.0871}\\ 
    \hline

  \multirow{3}{*}{6.M.A.}
  &Hf.&0.0904&0.0232&0.842&0.830&3192&0.6289\\
  &Db.&0.0722&0.0179&0.897&0.889&2027&0.2795\\
  &Eq.&\textbf{0.0648}&\textbf{0.0168}&\textbf{0.919}&\textbf{0.913}&\textbf{2016}&\textbf{0.2657}\\ 
    \bottomrule[1pt]
  \end{tabular}
}
\end{table}

\subsection{Main Results}
\label{exp:res}
In this subsection, we evaluate the proposed IceDiff with baseline methods for validation of the effectiveness.

\noindent\textbf{FM sets up new baselines for SIC prediction}  
IceDiff-FM is trained independently to predict SIC at 7 days, 8 weeks average, and 6 months average. As shown in Table~\ref{tab:main}, the performance of our proposed FM is superior to SICNet in all evaluation metrics. Although performance figures reported by other subseasonal and seasonal baseline methods may vary when using identical SIC datasets and training-validation-test split, our FM could provide competitive forecasts and sufficiently support down-scaling process.

\noindent\textbf{GDM achieves high down-scaling quality} 
We systematically compare the results of GDM
with other down-scaling methods.
In terms of interpolation-based methods, we select nearest, bilinear, and bicubic interpolation to compare with our diffusion-based GDM.
Besides, we select DGP~ \citep{pan2021exploiting} as a GAN-based method and GDP~\citep{fei2023generative} as a DDPM-based method for further comparison.
Interpolation-based methods simply upsample the low-resolution sea ice map, which uses existing pixel information to estimate missing pixel values.
DGP exploits the image prior captured by a GAN trained on large-scale natural images, while GDP utilizes a pre-train DDPM to achieve SOTA performance in blind image restoration.

As shown in \Cref{fig:high_res}, our GDM produces high-quality results 
and small-scale structures are successfully captured and rich details are generated through the guidance from the input map during each sampling step. 
By contrast, interpolation-based methods do not integrate details information of the low-resolution map, resulting in blurry edges and contours in the generated results. More quality analysis can be found in the appendix.

To further validate the superiority of our method, a metrics comparison is conducted between these four methods.
As shown in \Cref{GDM_metrics}, our method outperforms interpolation-based methods at the top of FID and Consistency rankings.
Compared to interpolation-based methods, our IceDiff-GDM improves FID score by 31.11 on average, reflecting it has the capacity to help generate a more realistic high-resolution map with a satisfactory texture restoration effect.
A lower Consistency demonstrates the structural difference between the down-scaled map and the original SIC input is relatively imperceptible, which better preserves the spatial distribution of sea ice forecast.

\subsection{Ablation Studies on Design of FM}
\label{exp:ablation}
As shown in Table~\ref{tab:abla.tsam}, by adopting TSAM to FM, the performance has improved for all metrics at 7 days and 6 months average lead times while the effects for 8 weeks average lead time are not all positive. Overall, incorporating TSAM to establish channel-wise dependency is favorable for SIC forecast models. 
To examine the impacts of input SIC data length, we trained FM for input lengths of half, double, and equal to the forecasting leads. The results in Table~\ref{tab:abla.io} show the peak of the performance at the ratio where the input length equals the forecasting leads.

\section{Conclusion}
\label{sec:con}
We propose a novel pan-Arctic SIC forecasting framework, IceDiff, which comprises IceDiff-FM and IceDiff-GDM, the former is capable of producing accurate forecasts at three different temporal scales and the latter leverages reliable forecasts as guidance to generate high-quality down-scaled SIC maps. The forecasting skills of IceDiff-FM at 7 days lead are superior to SOTA deep learning-based models. IceDiff-FM also sets up a new baseline for weekly average forecasting of 8 weeks and provides competitive results in seasonal SIC forecasts. To the best of our knowledge, IceDiff is the first attempt to forecast SIC at 6.25km spatial scale which is a quarter of the grid size (25km) that current methods performs on. The experimental results show that our framework can generate down-scaled SIC superior to off-the-shelf methods and it is vital for pushing forecasting models towards operational usage and further promotes sea ice researches.


\bibliographystyle{unsrtnat}
\bibliography{arxiv}

\end{document}